# Electrical Transport Properties of Liquid Pb-Li Alloys

S.G. Khambholja[1]* and A. Abhishek[2]

[1]Department of Science & Humanities, B & B Institute of Technology, Vallabh Vidyanagar, Gujarat, India

[2]Institute of Plasma Research, Gandhinagar, Gujarat, India

Corresponding author: physik.shyam@gmail.com

**Abstract**

It is generally observed that electrical transport properties of simple liquid metal based alloys can be explained well in terms of Faber-Ziman theory, $2k_F$ scattering model and finite mean free path approach. However, these approaches give poor description for materials, which show departure from nearly free electron (NFE) model. Taking Pb-Li as a test case of a system showing departure from NFE (which also exhibit compound formation tendency and disparate mass system), a new technique is proposed to compute electrical transport properties using model potential formalism coupled with t-matrix formulation. We have treated valence number of Pb & Li as a parameter in determining phase shifts. Further, rather than calculating phase shift in terms of Muffin-Tin potential, we have used model potential formalism. Present results suggest that compared to other three theoretical approaches mentioned above, present coupling scheme reproduce qualitative features of electrical transport properties of liquid Pb-Li alloys, which can be used further to study electrical transport properties of similar systems.

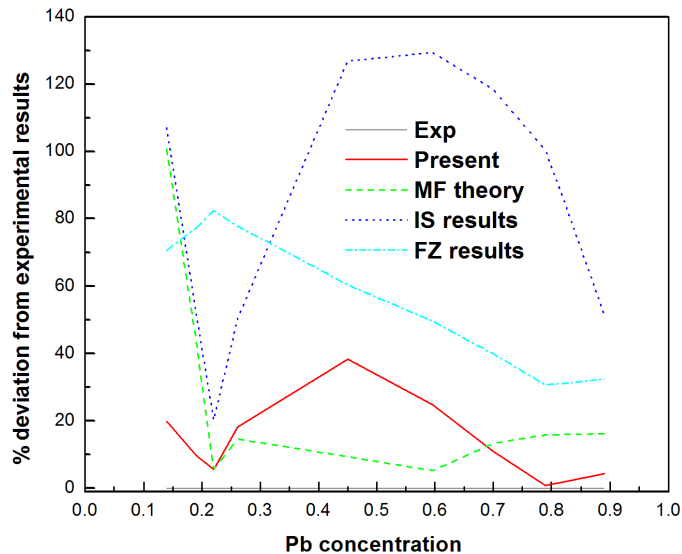

# 1. Introduction

Lead-Lithium (Pb-Li) alloys in its eutectic composition ($Pb_{83}Li_{17}$) are considered as one of the important candidates among materials being tested as liquid blankets [1-3]. At international level, large amount of experimental work is carried out to understand various properties of Pb-Li under high temperature conditions [4-10] and various thermoelectric potentials are developed [11, 12]. On the other hand, less theoretical attempts are performed so far to compute various physical and chemical properties of Pb-Li alloys [13-19]. Model potential formalism is able reproduce correct trends in properties of interest with less computational efforts. It has been used earlier by many researchers to study various physical properties of liquid metals and alloys [20-22]. Thus, in the present work, we have used model potential formalism to incorporate ion-electron interaction.

Literature survey suggest that around eutectic composition ($Pb_{83}Li_{17}$), anomalies are observed in electrical transport properties and thermodynamic properties [23] etc, which might be associated with compound formation tendency [24]. It is noted by Hoshino et al [25] that the effective number of conduction electrons has its lowest value around eutectic composition. These authors have calculated electrical resistivity of Pb-Li liquid alloys using "molecular formation" theory. Pastruel et al [26] have calculated electrical resistivity of Pb-Li alloys using quantum interference approach, which is basically a modification of Faber-Ziman (FZ) theory. As noted by Hafner et al [27], simple FZ approach has some limitations in determining electrical transport properties of liquid alloys. The reason behind such limitations may be effects of chemical ordering as reflected in structure factors or strong incoherent scattering. In their experimental determination of electrical resistivity of Li based alloys, Nguyen et al [24] have noted that for $Z \geq 3$, the values of electrical resistivity are far away from nearly free electron theory. Thus, it is expected that the FZ approach or modified FZ approach may give results which are in poor agreement with the experiments. However, an attempt was made by Khajil et al [28] who have used non local pseudopotential to compute electrical resistivity within FZ model. The results were highly underestimated. Hoshino et al [25] have noted that calculation of electrical resistivity within relaxation time approximation may not be valid due to its high electrical resistivity around eutectic composition. Meijer et al [29] have reported experimental results of electrical resistivity of liquid Pb-Li alloys along with its temperature variation and have noted that near eutectic composition, Pb-Li alloys exceeds the limiting value for metallic behavior. They have attributed this feature to the known fact that in liquid Pb-Li alloys; nearly free electron approximation may not be valid in some concentration range.Thus in this situation where almost all theoretical approaches provide poor description of electrical transport properties of liquid Pb-Li alloys, we thought it worthwhile to compute electrical resistivity of liquid Pb-Li alloys using t-matrix formulationin conjunction with model potential formalism.The necessary modification based on density of states is incorporated in the present calculations. The paper is organized as following. Next section 2 deals with the theory and method of computation. Section 3 deals with the results and discussion part and the paper are concluded in section 4.

## 2. Method of Computation

### 2.1 Electrical resistivity

In the t-matrix formulation, electrical resistivity of liquid binary alloys is calculated using following equation [30-32].

$$\rho_{el} = \frac{\pi \cdot \Omega_0}{k_F^2} \int_0^1 \left(\frac{q}{2k_F}\right)^3 d\left(\frac{q}{2k_F}\right) |T_{alloy}|^2 \tag{1}$$

Here, $|T_{alloy}|^2$ is the total t-matrix form factor and is given as following.

$$|T_{alloy}|^2 = c_i.|t_i|^2 \left[1-c_i+c_i a_{ii}(q)\right] + c_j.|t_j|^2 \left[1-c_j+c_j a_{jj}(q)\right] + c_i.c_j \left(t_i^* t_j + t_i t_j^*\right)\left[a_{ij}(q)-1\right] \tag{2}$$

Here, $t_i$ and $t_j$ are the t-matrix form factors of pure element forming alloys [in the present case Pb and Li]. $a_{ii}(q)$, $a_{jj}(q)$ and $a_{ij}(q)$ are the Ashcroft-Langreth partial structure factors [33] corresponds to Pb-Pb, Li-Li and Pb-Li, respectively. $c_i$ is the concentration of Pb and $c_j$ is the concentration of Li such that $c_i + c_j = 1$. In determining the t-matrix element for the alloys, we assumed that phase shifts of pure Pb and Li are independent of alloy concentration. t-matrix form factors of pure elements are calculated using following equation.

$$t(E_F, q) = -\frac{2\pi}{\Omega_0 (2E_F)^{1/2}} \sum_l (2l+1) \sin\eta_l \exp(i.\eta_l) P_l(\cos\theta) \tag{3}$$

Phase shifts $\eta_l$ are calculated up to second order. Generally, while calculating electrical resistivity within t-matrix formulation, Muffin-Tin (MT) potential is used to determine phase shifts [31]. Dreirach et al [35] have proposed to compute Fermi energy from zero of MT potential. In another approach of Esposito et al [36], the authors have determined Fermi energy by neglecting $E_b$ term. The authors in Ref. [32] have determined Fermi energy by introducing effective valence as a parameter. In the present work, we propose a slight alternative approach. Present work is based on following modification.

(i) We have taken $E_b = 0$.

(ii) Rather than calculating phase shifts in terms of MT potential, we have calculated them using model potential formalism. This modification allows us to incorporate different types of model potential in our calculation.

(iii) Further, we have treated effective valence in pure Pb and Li as a parameter.

We have used valence as parameter to fit the presently calculated electrical resistivity of pure Pb and Li with the experimental values. This parameterization is based on the observation that in Pb, experimental and theoretical results of density of states predict $Z \approx 2$ [38], whereas the accepted value is 4. This approach is in a sense an alternative approach in t-matrix formulation, the validity of which shall be discussed later on. Further, we have not performed any other kind of fitting procedure in our calculation.

In determining phase shifts, we have used Ashcroft empty core model (ECM) potential [39] to incorporate ion-electron interaction in conjunction with Percus-Yevick (PY) hard sphere structure factors [40]. ECM is a single parametric model potential and has been used by many authors to calculate various properties of metals and alloys. The only parameter of this model is core radius, which is determined from the known density of Pb and Li [22]. ECM potential is given as following.

$$V_b(q) = -\frac{4 \pi z e^2 \cos(q r_c)}{\Omega_0 q^2} \qquad (4)$$

Experimental values of densities of both Pb and Li are adopted from literature [40]. Local field correction function due to Ichimaru and Utsumi(IU) [41] is used to calculate screened form factors of Pb and Li. This function satisfies compressibility sum rule in the long wavelength limit and is used over wide range of metallic densities. Three parameters of IU screening functions are determined in terms of known densities of Pb and Li. Once the t-matrix form factors of pure Pb and Li are obtained, they are used along with Ashcroft-Langreth partial structure factors in equation (2) and finally equation (1) is used to compute electrical resistivity as a function of Li concentration.

## 2.2 Thermal Conductivity

Based on the knowledge of electrical resistivity, thermal conductivity of Pb-Li alloys is calculated using following expression [42].

$$\sigma = \frac{\pi^2 k_B^2 T}{3|e|^2 \rho_{el}} \qquad (5)$$

Above expression (6) is used based on the argument that coefficient of thermal conductivity does not show a marked gap at solid-liquid phase boundary. In expression (6), $k_B$ is Boltzmann's constant.

## 3. Results and Discussion

Presently calculated t-matrix form factors of pure liquid Pb and Li are shown in Fig. 1. Fig. 2 shows presently computed phase shifts for pure Pb and Li. Since computation of phase shifts depends on particular MT potential, no data for comparison are available. It is observed that zeroth order phase shift (s-phase shift) contribute maximum at lower energy and at higher energy values, contribution is more from first and second order phase shifts. Presently calculated energy dependent electrical resistivity of pure Pb and Li are shown in Fig. 3 along with the results of Abdellah [32]. In their approach, Abdellah et al have determined Fermi energy with respect to MT zero potential and have taken $E_B \neq 0$. Further, they have used effective valence as a parameter in their calculation. In the same paper [32], other two approaches namely one due to Dreirach et al [35] and another due to Esposito et al [36] are also cited. Dreirach et al uses $E_B \neq 0$ and determines effective mass from band structure of crystalline state. On the other hand, approach of Esposito et al uses $E_B = 0$ and determines Fermi energy by filling density of states curves by Z electrons. Further, using all three approaches, Abdellah et al [32] have reported electronic properties of liquid lead using two different values of Z, Z = 2 and Z = 4. Unlike these approaches, we have taken $E_B = 0$ and have treated Z as a parameter to obtain experimental resistivity of pure Pb and Li. We obtain Z = 1.75 for Lead and Z = 0.58 for Lithium from our calculation. Another reason for treating valence as a parameter is based on the following observations. Altshuler et al [43] have noted an anomaly in the density of states near Fermi surface in case of non-crystalline materials. This anomaly is associated with the reduction of effective valence. Electrical conductivity is proportional to density of state near Fermi surface. But, in the regions where Edward's cancellation rule can not be not applied (as resistivity values are more compared to limiting values for metallic behavior), conductivity is proportional to normalized density of states i.e. $N(E_F)/N(E_F)_{free}$ as noted by Meijer et al [29]. Due to this observation, effective valence as well as Fermi wave number are greatly affected and reduce from ideal values. In their work, Meijer et al have noted that Pb-Li system shows departure from nearly free electron model. Thus, it is expected that effective valence of alloy may vary from ideal values in such situations. The low energy deviation of presently calculated electrical resistivity from that of Abdellah et al [32] may be due to difference in approaches. However, at higher energies around equilibrium, presently calculated values are in line with the experimental results. The difference between these values may be attributed to the arbitrariness of model potential inside the core. Further, we have determined phase shifts from model potential calculations and not from MT potential. Presently calculated values of Fermi energy and electrical resistivity of pure lead are reported in Table 1 along with the results mentioned in Ref. [32]. It is seen that presently calculated values are falling in between those reported in Ref. [32]. It is also observed that even when Dreirach approach includes band structure energy term, corresponding values of electrical resistivity are much higher compared to experimental results.

We have calculated Ashcroft-Langreth partial structure factors of Pb-Li alloys in the whole concentration range. These structure factors are reported in our recent communications [13,14] and hence are not plotted here. Presently calculated electrical resistivity of liquid Pb-Li alloys at 873K temperature are shown in Fig. 4 along with experimental and other theoretical results. In low and high Pb concentration region, presently calculated values match nicely with the experimental results. The peak position in resistivity is also correctly reproduced. At intermediate Pb concentration, presently calculated values are slightly overestimated. We have also plotted the FZ theory based results of Khajil et al [28]. Limitations of FZ theory areclearly seen in case of alloys like Pb-Li. On the other hand, finite mean free path based approach [28] overestimate the resistivity values. Results based on molecular formation theory are higher at low Li concentrations [25]. In a similar way, results reported by Pasturel et al [26] are also higher compared to experimental results. They have used using $2k_F$ scattering model (these results are not plotted here to avoid confusion in the graph). Meijer et al [29] have noted that electrical resistivity values of alkali-lead alloys exceed the limiting value of metallic behavior. This feature is correctly reproduced in the present work. Since the peak position in the resistivity isotherm is also correctly reproduced, we believe that compared to diffraction model based FZ formulation, present coupling scheme is incorporating N(E) near Fermi surface in a far better way and hence present results are better.

Table 1. Presently computed values of Fermi energy and electrical resistivity of pure lead are compared with those mentioned in Ref. [32].

| Name of approach | Lead | Remarks |
| --- | --- | --- |
| Abdellah et al [32] | $E_B$ = -0.24 Ry<br>$E_F$ = 0.2534 Ry<br>$\rho_{el}$ = 148.11 $\mu\Omega.cm$<br>T = 876 K | Z = 2 |
| Esposito et al [36] | $E_B$ = 0 Ry<br>$E_F$ = 0.47 Ry<br>$\rho_{el}$ = 68.80 $\mu\Omega.cm$<br>T = 876 K | Z = 2 |
| Dreirach et al [35] | $E_B$ = -0.24 Ry<br>$E_F$ = 0.1655 Ry<br>$\rho_{el}$ = 435.74 $\mu\Omega.cm$<br>T = 876 K | Z = 2 |
| Present results | $E_B$ = 0 Ry<br>$E_F$ = 0.376 Ry<br>$\rho_{el}$ = 106 $\mu\Omega.cm$<br>T = 873 K | Z = 1.75 |

Presently calculated values of thermal conductivity are plotted in Fig. 5 along with the experimental results at 623K temperature [42]. Present calculations are performed at 873K temperature and experimental results for Pb-Li eutectic ($Pb_{83}Li_{17}$) are available up to 623K only. Thus, a clear difference is seen between both of these results.

## 4. Conclusion

In the present work, we have proposed a new technique for the calculation of electrical resistivity of binary alloys like Pb-Li, which show departure from NFE model. We have used model potential formalism to compute finite phase shifts, which are further used to compute t-matrix form factors. Present coupling scheme does not include any parameterization expect fitting experimental resistivity of pure elements, which has a proper justification based on experimental and theoretical density of states curves. Still it reproduces experimental results in a far better way compared to FZ model, $2k_F$ scattering model and finite mean free path model. This approach can be used to compute electrical resistivity of similar alloys, which will be helpful to understand density of states, departure from NFE and Mott transitions, if any and fluid MHD regimes.

**Acknowledgement**

Authors are grateful to Board of Research in Nuclear Science, Government of India for financial assistance under a Major Research Project (39/14/29/2016-BRNS).

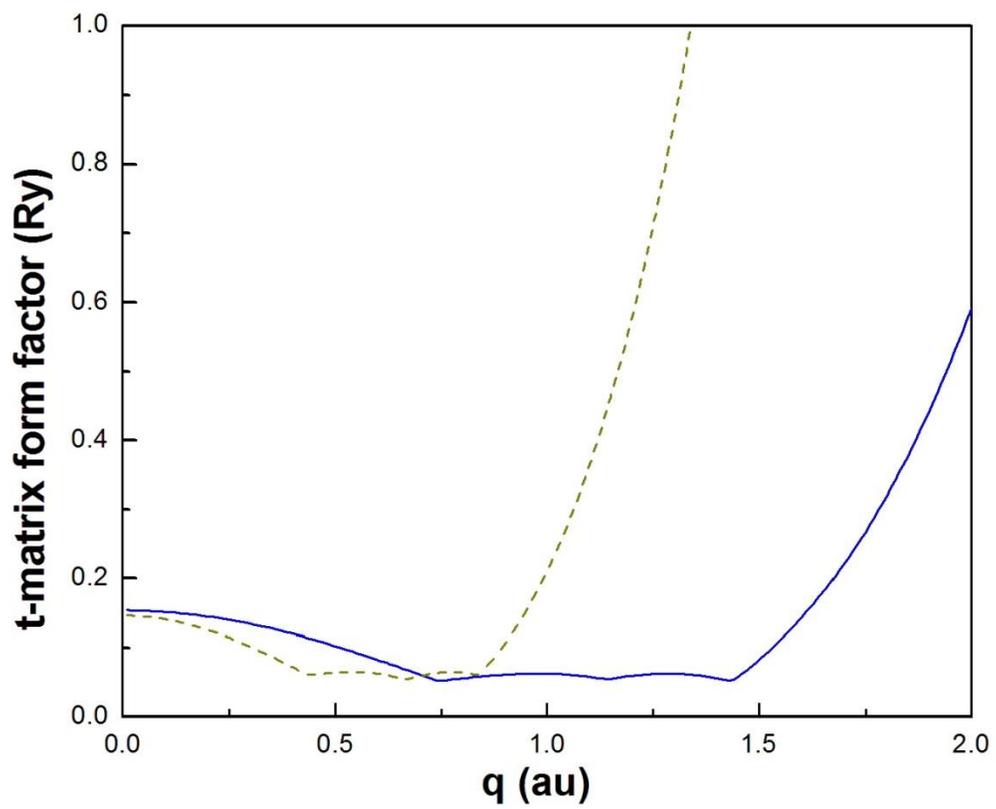

Figure 1. t-matrix form factors of pure liquid Pb (full line) and Li (dashed line).

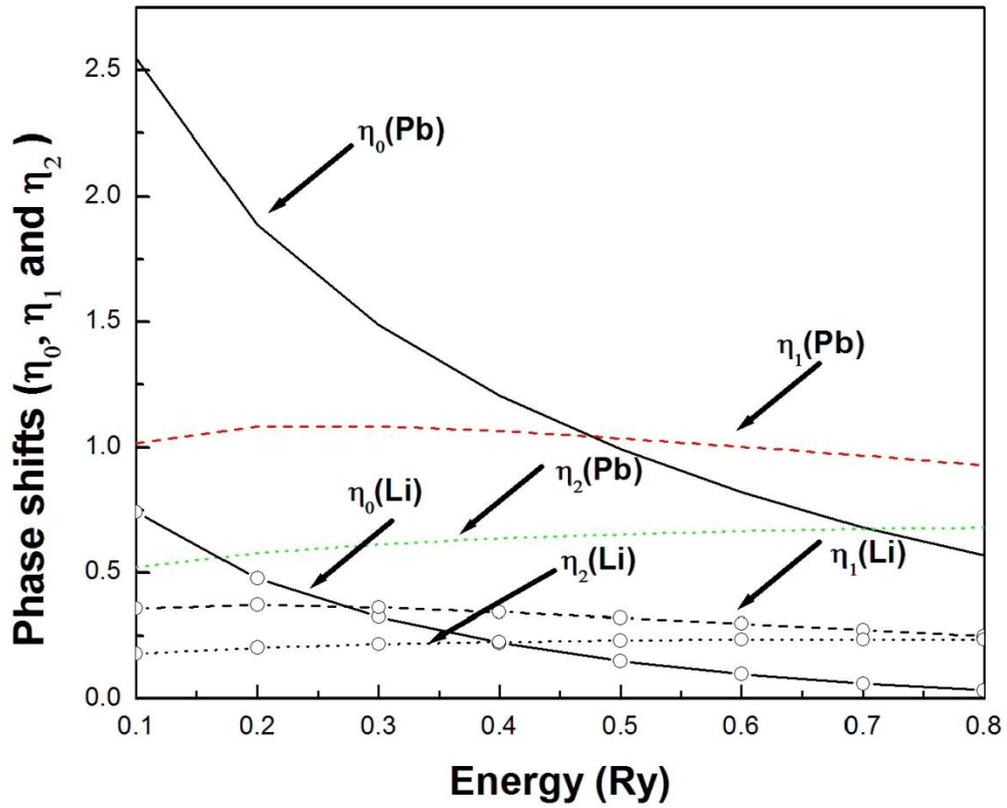

Figure 2. Presently computed phase shifts of Pb and Li.

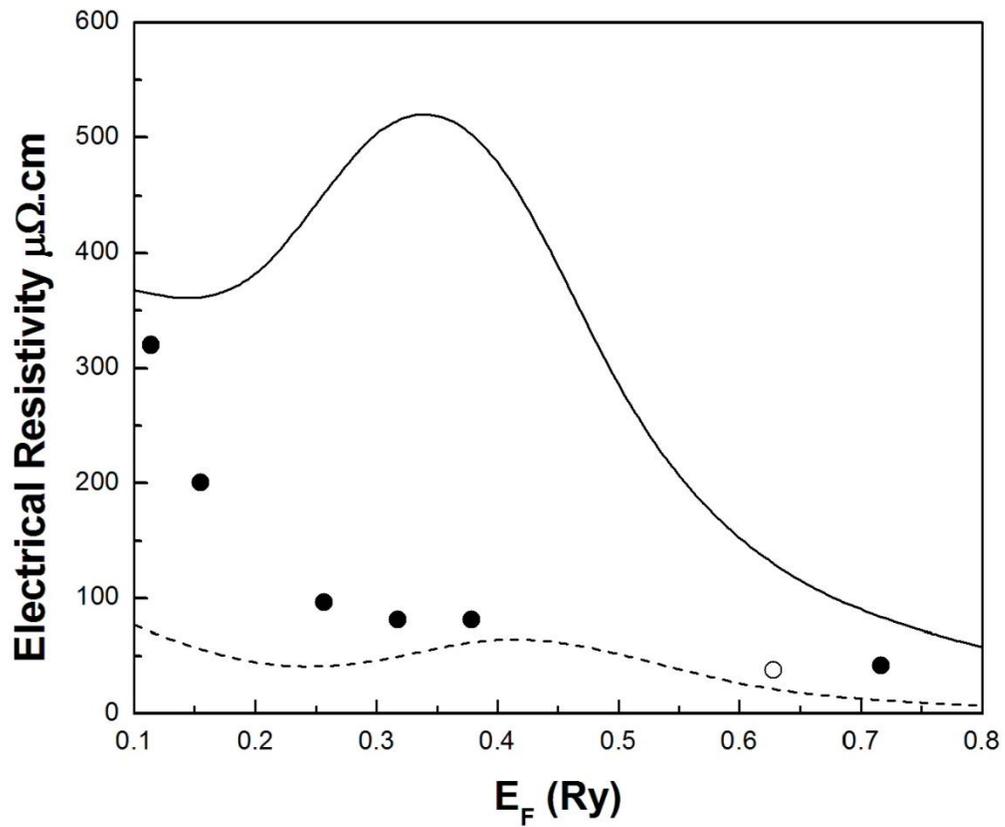

Figure 3. Presently calculated energy dependent electrical resistivity of liquid lead (full line) along with the results of Abdellahat el (filled and open circles) [32] at 750 °C temperature. Dashed line is the presently calculated energy dependent resistivity of liquid lithium.

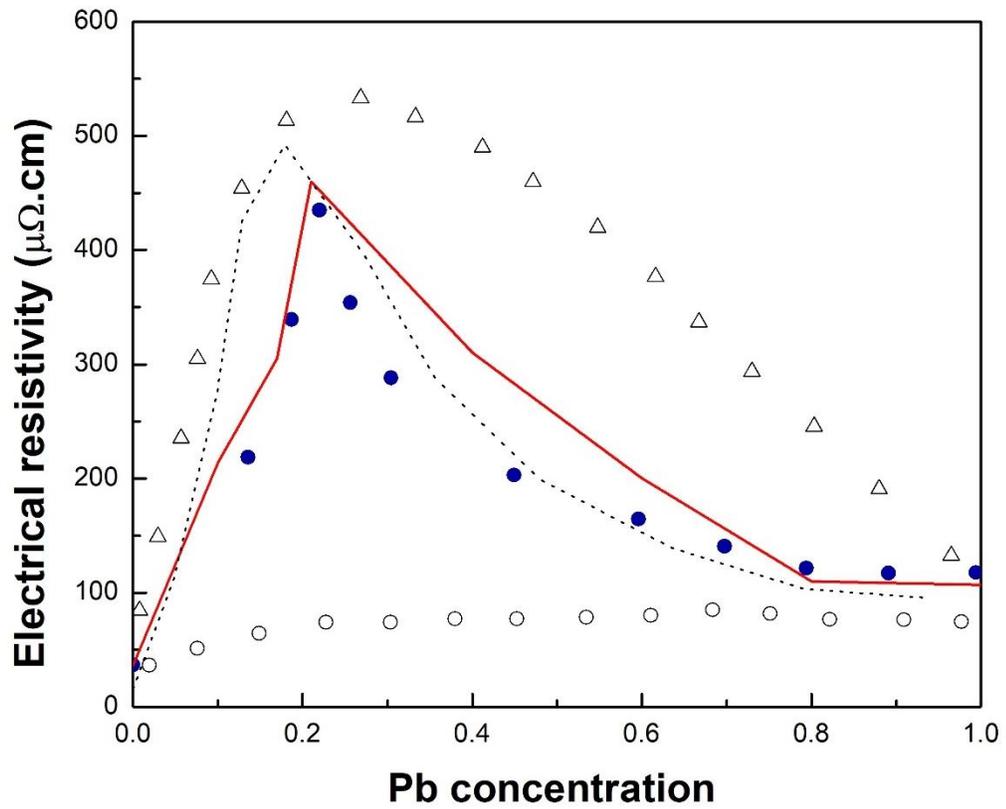

Figure 4. Presently computed electrical resistivity (full line) along with experimental results (filled circles) [29], theoretical results of molecular formation theory (dashed line) [25], FZ results of Khajil et al (open circles) [28] and IS results of Khajil et al (open triangles) [28].

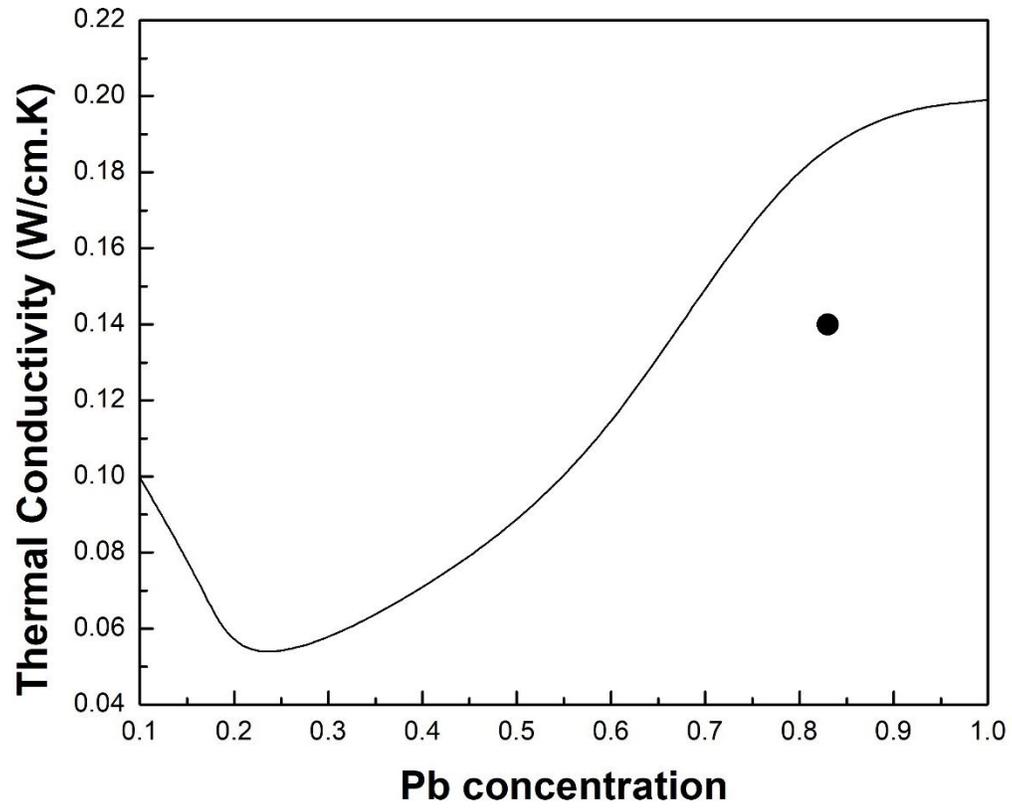

Figure 5. Concentration dependent thermal conductivity of liquid Pb-Li alloys (full line) at 873K along with experimental results (symbol) [42] at 623K.